# Vacuum Formed Temporary Spherical and Toroidal Bent Crystal Analyzers for High Resolution X-ray Spectroscopy

Evan P. Jahrman[1] (§), William M. Holden[1] (§), Alexander S. Ditter[1,2], Stosh A. Kozimor[2], Scott L. Kihara[1], Gerald T. Seidler[1] (*).

[1]Physics Department, University of Washington, Seattle WA 98195
[2]Chemistry Division, Los Alamos National Laboratory, Los Alamos, NM 87545

We demonstrate that vacuum forming of 10-cm diameter silicon wafers of various crystallographic orientations under an x-ray permeable, flexible window can easily generate spherically bent crystal analyzers (SBCA) and toroidally bent crystal analyzers (TBCA) with ~1-eV energy resolution and a 1-m major radius of curvature. In applications at synchrotron light sources, x-ray free electron lasers, and laboratory spectrometers these characteristics are generally sufficient for many x-ray absorption fine structure (XAFS), x-ray emission spectroscopy (XES), and resonant inelastic x-ray scattering (RIXS) applications in the chemical sciences. Unlike existing optics manufacturing methods using epoxy or anodic bonding, vacuum forming without adhesive is temporary in the sense that the bent wafer can be removed when vacuum is released and exchanged for a different orientation wafer. Therefore, the combination of an x-ray compatible vacuum-forming chamber, a library of thin wafers, and a small number of forms having different secondary curvatures can give extreme flexibility in spectrometer energy range. As proof of this method we determine the energy resolution and reflectivity for several such vacuum-formed bent crystal analyzers (VF-BCA) in laboratory based XAFS and XES studies using a conventional x-ray tube. For completeness we also show x-ray images collected on the detector plane to characterize the resulting focal spots and optical aberrations.

(*) seidler@uw.edu
(§) These authors contributed equally to this work.

## I. Introduction

Doubly-curved Bragg optics see extensive use in advanced x-ray spectroscopies at x-ray free electron lasers, synchrotron x-ray light sources, laser plasma facilities, and laboratories performing in-house x-ray absorption fine structure and x-ray emission spectroscopy. This has been made possible by a large body of work aimed at characterizing and evaluating the properties of doubly-curved optics.[1-7] These optics are available commercially; the traditional production methods use glue or anodic bonding to affix crystal wafers after pressing into precision glass or metallic substrates.[8-11] Other production techniques such as vacuum-bent analyzers[12] and hot plastic deformation techniques[13] have also been explored. Another approach is to use a spherical bending apparatus that allows the curvature to be characterized before fixing to a permanent shape.[14]

Even with a variety of available production techniques, acquiring a large number of such optics for multi-analyzer systems can be challenging. Synchrotron endstations for x-ray emission spectroscopy (XES), high energy resolution fluorescence detection (HERFD) of x-ray absorption fine structure (XAFS), or nonresonant inelastic x-ray scattering (NRIXS) now often use from five to as many as ~100 such optics.[15-22] In addition to cost issues, for XES there is a pragmatic issue: different synchrotron users may come with experiments requiring acquisition of an entirely new set of optics for some energy range that has not previously been explored at that facility. The same issue arises with the increasing use of laboratory-based spectrometers,[23-31] where again the elements and energy ranges that are capable of being studied depend on the availability of analyzers having a variety of crystal cuts to achieve the necessary Bragg angles and energy resolutions. As an additional issue, when used in a Rowland circle spectrometer the most common doubly-curved optics, spherically bent crystal analyzers (SBCA) in the Johann configuration[32], sagittal bending error results in large spot sizes out of the Rowland plane when far from backscatter. Toroidally bent crystal analyzers (TBCA) are needed for a truer point-to-point focus, but are far less common and their secondary curvature must be tuned to match a chosen Bragg angle.

The purpose of this paper is to address the above concerns, at least for applications requiring only the 'modest' energy resolution of ~1-eV, such as is frequently sufficient in the chemical sciences.[33] Specifically, we demonstrate a simple method to make temporary SBCA's and TBCA's with good performance and very high ease of use. Unlike methods using epoxy or anodic bonding, vacuum forming without adhesive is temporary in the sense that the bent wafer can be removed when vacuum is released and exchanged for a different orientation wafer. The combination of an x-ray compatible vacuum-forming chamber, a suitable library of thin single-crystal wafers, and a small number of forms having different secondary curvatures can give extreme flexibility in spectrometer or monochromator energy range at low cost and with great flexibility for addition of new energy ranges, as needed.

## II. Methods

The overall concept and resulting design of the vacuum-formed bent crystal analyzer (VF-BCA) is presented in Fig. 1. Front-view and section-view computer aided design (CAD) renderings are shown in Fig. 1(a) and (b), respectively. The key point is that an x-ray transparent, flexible polyimide window serves to seal the volume containing the wafer and the machined aluminum alloy or glass lens form from the outside atmosphere. When pumping on the VF-BCA, outside air pressure forces the polyimide window to collapse inward, pressing the wafer firmly onto the underlying form. A photograph of a VF-BCA is shown in Fig. 1(c), and for comparison, a commercial, anodically-bonded analyzer (XRS TECH LLC) is shown in Fig. 1(d).

Several different supporting forms were used, all with the 1-m major radius of curvature required by the laboratory spectrometer[27] that served as an extremely convenient testbed for this project. Table 1 defines the character of the different forms. When the forming contact surface was machined from aluminum 6061 alloy, a Mitsubishi M-V5Cn-L vertical machining center was used. The necessary cutting paths were generated under the target scallop deviation of 5 $\mu$m with a ball end mill having a ball-end radius of ~7.14 mm. The resulting surfaces were cleaned with solvents but otherwise not modified by, e.g., polishing or lapping.

The wafers used in the VF-BCA were standard 10-cm diameter double side polished Si wafers of various orientations, all nominally 400-$\mu$m thick. Different crystal orientations yield different energy ranges over the useful Bragg angle range. A list of absorption edges or fluorescence lines studied and the corresponding commercial SBCA or wafer used in a VF-BCA is given in Table 2.

The laboratory spectrometer used here[27] is a 1-m diameter Rowland circle spectrometer based on the approach of Seidler, et al.,[23] as modified by tilt-free alignment[34] and by the use of a higher-powered x-ray tube and longer translation stages to give a wider range in Bragg angle ($\theta_B$). Across several instrument generations this overall approach using a conventional x-ray tube and a 'scissors' monochromator has been used in several studies with ~1-eV energy resolution for either XAFS or XES[24, 26, 29, 34-36] and the present instrument and its performance have been described in detail in Jahrman, et al.[27]

Measurements were performed with a Varex VF-80 x-ray tube with Pd-anode operating at 35 kV accelerating potential and 100 W total electron beam power. A silicon drift diode (Amptek X-123 SDD) with ~4.6-mm diameter active region was the final detector for all XAFS and XES scans. In Fig. 2 we show a schematic of the Rowland circle implementation for this spectrometer, a photograph of the spectrometer in an XAFS configuration, and a photograph of the VF-BCA installed at the optic location.

Under the protocol for tilt-free alignment,[34] spherical analyzers are rotated about their circular symmetry axis until the wafer's miscut is in the Rowland plane. For a toroidal VF-BCA, the wafer orientation must be determined in a spherical VF-BCA prior to installation into the toroidal vacuum form holder with the miscut oriented in the Rowland plane, i.e., in the plane of the 1-m major radius of curvature. The use of a sliding magnet mount (Fig. 2(c)) makes it particularly convenient to rotate about the necessary axis in the Rowland plane.

Imaging of the x-ray intensity on the detector plane was performed using a small home-built CMOS x-ray camera. This is an updated version[37] of an earlier camera[38] that has seen good use in a lower-energy XES instrument.[39-42] The camera has a $3.2 \times 5.6$ mm$^2$ field of view. It was mounted on a micrometer-driven vertical translation stage and manually repositioned to achieve mosaic coverage of the x-ray intensity's spatial distribution. Furthermore, the camera's ability to identify both the location and energy of individual x-ray photons allowed the rejection of stray fluorescence by energy-windowing.

Ray tracing software written in `Mathematica` was used to assess Johann error and beam spreading perpendicular to the Rowland plane due to sagittal error, i.e., the use of optics whose second radius of curvature is not equal to the perpendicular distance from the optic center to the line connecting the source and detector points on the Rowland circle (the sagitta). Unlike recent work aiming to give an advanced treatment of the interplay between strain effects and, e.g., dynamical diffraction in SBCAs,[43,44] here we only seek purely geometric optics effects on slightly sub-mm length scales. Consequently, the Monte-Carlo ray tracing code generated x-rays from a 1-mm diameter source spot, reflected them from the bent optic using simple, specular Bragg reflection assuming zero wafer miscut, and then recorded the position of the intersection of those rays with the detector plane.

XAFS measurements were performed on a 6-µm thick Ni foil from EXAFS Materials. XES measurements were performed on a 75-µm thick sheet of commercial Cu foil. All measurements were performed with the sample in air under ambient conditions, and a helium space was used to reduce air-absorption. In some cases, small corrections for slow leaks in the helium space have been made to ensure that all comparisons are on a common efficiency basis. All XAFS spectra were dead time corrected and subsequently processed in Athena where standard background removal and normalization procedures were followed.[45] For XES, all spectra are dead time corrected and approximately aligned to a common energy scale.

## III. Results

Optic performance encompasses both its focal properties, as this is crucial for coupling to the final detector, and also its energy resolution. We begin with focal properties. In Fig. 3 we present the x-ray intensity in the detector plane for the commercial, anodically bonded Si (551) SBCA and for a Si (711) wafer in VF-BCA-1, VF-BCA-2, and VF-BCA-3 (see Table 1 for the definitions of these terms). Note that these two crystal orientations have the same *d*-spacing, and consequently are identical for present purposes, giving the same energy range over the same span of Bragg angles. The qualitative agreement is very good. All optics show strong horizontal focusing and also the expected degree of vertical focusing subject to sagittal distortion. VF-BCA-2 and VF-BCA-3, which are based on machined, unpolished metal forms, have slightly inferior focal properties. As discussed below, precision polishing of machined surfaces is an obvious future direction for improvement.

In Fig. 4(a), we show the horizontal intensity spread across the detector plane at different Bragg angles for each of the above optics and a Si (711) or equivalent wafer. Although the in-plane focal qualities of SBCA and VF-BCA-1 are similar, the horizontal profile of VF-BCA-2 is found to be broader and skewed. The vertical intensity spread is shown in Fig. 4(b). Near backscatter, the out-of-plane focal quality of both VF-BCA-1 and VF-BCA-2 is found to be comparable to the SBCA. At lower values of $\theta_B$, VF-BCA-1 and VF-BCA-2 demonstrate clear inhomogeneities, although the total refocused intensity remains comparable.

For the out-of-plane focal quality, it is clear that the use of a SBCA sufficiently far away from backscatter ($\theta_B = 90$ deg) results in rapid spreading of the beam in the out-of-plane direction as expected from sagittal error. In this configuration, the vertical spread of the beam exceeds the height of the detector's active area, as shown in Fig. 4(b). This raises the question of using toroidal optics where the primary radius of curvature is dictated by the Rowland circle diameter but where the secondary radius of curvature is chosen for ideal point-to-point focusing for a selected 'design' Bragg angle $\theta^*$. Ray-tracing calculations for the out-of-plane beam height as a function of $\theta_B$ are shown in Fig. 5(a) for TBCA's having design $\theta^*$ varying from 55 to 90°, the lattermost being simply an SBCA. These simulations strongly suggest that TBCA should give a more efficient coupling to the finite-sized detector when the secondary radius of the TBCA is chosen to eliminate sagittal error for $\theta^*$ in the middle of the angular range dictated by the energies of interest. Consequently, in the bottom panel of Fig. 3 we show the intensity distributions on the detector plane for VF-BCA-3 with a Si (711) wafer. The out-of-plane focal properties of the TBCA are clearly much superior to the SBCA when $\theta_B$ is in the vicinity of the designed $\theta^*$.

The improved spectrometer performance when using a TBCA at $\theta_B$ near the designed $\theta^*$ is demonstrated in Fig. 6. Here, the intensity of x-rays refocused at the SDD by each optic is shown across the full angular range of the instrument. The very short detectors used in the test studies (~4.6-mm active height) gives an especially high sensitivity to vertical beam spread, resulting in the narrow experimental Bragg angle range for optimum performance of the TBCA. Ray tracing calculations for TBCA simulating different detector heights are given in Fig. 5(b). In each case, there is an optimal, flat-top region of Bragg angles where all of the reflected x-rays are collected by the detector when the height of the reflected beam is smaller than the detector diameter. This agrees well with the experimental data of Fig. 6, which shows the same flat top near the $\theta^*$, and a decrease in count rate far from $\theta^*$. Further, the ray tracing demonstrates the utility of a larger detector which increases both the width of the flat-top region as well as the count rate when $\theta_B$ is far from $\theta^*$.

The preceding discussion has only addressed focal properties. Now we report on the energy resolution using the vacuum clamped optics. The measured Ni K-edge XANES for all optics using a Si (551) or equivalent wafer are shown in Fig. 7(a). It can be seen that all optics produced nearly identical spectra, suggesting a negligible loss in energy resolution from the commercial to vacuum clamped optics. The high quality of the XANES spectra is typical of modern laboratory based XAFS systems.[23, 24, 27, 46, 47] Similarly, Fig. 7(b) presents again Ni K-edge XANES where the spectra are instead measured with the Si (444) reflection by several different optics. Again, spectra are nearly identical, with only minor differences observable in the extent of the shoulder at ~8334 eV and in the magnitude of the oscillation at ~8352 eV

In Fig. 7(c), Cu Kα XES results are presented for all optics using the Si (444) reflection. Here again, spectra were found to be nearly identical, however some small differences in peak ratios can be observed in accordance with small differences in energy resolution and also small differences in the angular response functions between optics. Recall again that the analyzed radiation is being imperfectly focused in the vertical, with a spread larger than the detector height. Consequently, small changes in spectrometer alignment can lead to few-percent differences in net monochromator efficiency as a function of energy.

Given the success and limitations shown in this effort to make temporary doubly-bent crystal analyzers, there are several future directions that merit comment. First, the focal quality of optics should be improved by lapping and polishing the surfaces of the machined forms, or by acquiring precision surface-ground glass forms for the toroidal case, in analogy to the high-quality lens used for the spherical case. While this is not particularly relevant for the ~1-eV resolution needed for many measurements in a point-to-point Rowland circle configuration, the same would not be true for higher-energy resolution applications or the important case of dispersive spectrometers based on spherical analyzers, as is commonly used at synchrotron light sources.[48-50] Second, the efficiency of each optic across wider angular ranges of the instrument could be improved by implementing a taller detector. Third, it would be interesting to explore forms with smaller primary radii of curvature or using wafers composed of crystalline materials besides silicon, each with the goal of obtaining higher signal levels. Fourth, although we have only used this method for intact round wafers, one should expect that the same apparatus can be used for segmented wafers, such as is used in the recent development of 0.5-m radius of curvature SBCAs,[9] or for pieces of multiple wafers integrated to obtain a larger BCA solid angle than could be obtained with any single wafer. This lattermost opportunity is likely relevant for materials where the *de facto* standard 10-cm diameter wafers are not available.

## IV. Conclusions

We report the development and performance of spherically and toroidally bent crystal analyzers for use in high resolution x-ray spectroscopies. Unlike the present practice of gluing or bonding the necessary crystalline wafers to a high-precision glass lens, we instead use only air pressure to hold the wafer against the shaping form that provides the necessary profile. The specifications for the shaping form are found to be rather modest, in that modern machined metal forms suffice and high-precision, high-cost surface ground lenses are not needed. The resulting optics demonstrate resolutions and efficiencies comparable to their commercially available counterparts as determined by XANES and XES measurements using a laboratory spectrometer. These results establish a considerable convenience, simplicity, and flexibility that may prove useful for Rowland circle spectrometers in the lab for XAFS and XES studies, as well as at synchrotron and x-ray free electron laser x-ray facilities for XES, high-energy resolution fluorescence detection (HERFD), and resonant inelastic x-ray scattering (RIXS).

## Acknowledgements

E. Jahrman was supported in part by the Joint Center for Energy Storage Research (JCESR), an Energy Innovation Hub funded by the U.S. Department of Energy, Office of Science, and Basic Energy Sciences, and by the U.S. Department of Energy through the Chemical Science and Engineering Division of Argonne National Laboratory. W. Holden and G. Seidler were supported by the Joint Plasma Physics

Program of the National Science Foundation and the Department of Energy under Grant No. DE-SC0016251.  R&D associated with the Los Alamos National Laboratory (LANL) spectrometer was funded under the Heavy Element Chemistry Program by the Division of Chemical Sciences, Geosciences, and Biosciences, Office of Basic Energy Sciences, U.S. Department of Energy and the U.S. Department of Energy. LANL is operated by Los Alamos National Security, LLC, for the National Nuclear Security Administration of U.S. Department of Energy (contract DE-AC52-06NA25396).

| Optic name | Major radius (cm) | Perpendicular radius (cm) | Vacuum form surface |
|---|---|---|---|
| SBCA | 100.0 | 100.0 | Wafer anodically bonded to glass |
| VF-BCA-1 | 100.0 | 100.0 | Glass lens |
| VF-BCA-2 | 100.0 | 100.0 | Al6061 spherical recess |
| VF-BCA-3 | 100.0 | 88.3 | Al6061 torus, $\theta^* = 70°$ |

**Table 1**: List of all bent crystal analyzers and analyzer forms used in this study.

| Study | Commercial SBCA | Wafer for VF-BCA | Bragg angle (deg) |
|---|---|---|---|
| Ni XAFS | Si (551) | Si (711) | 78.0 |
| Ni XAFS | Si (444) | Si (444) | 71.6 |
| Cu Kα XES | Si (444) | Si (444) | 79.3 |

**Table 2**: List of experiments performed, commercial SBCA or wafers used in the VF-BCA, and nominal Bragg angle for the absorption edge or fluorescence line from the indicated crystal reflection.

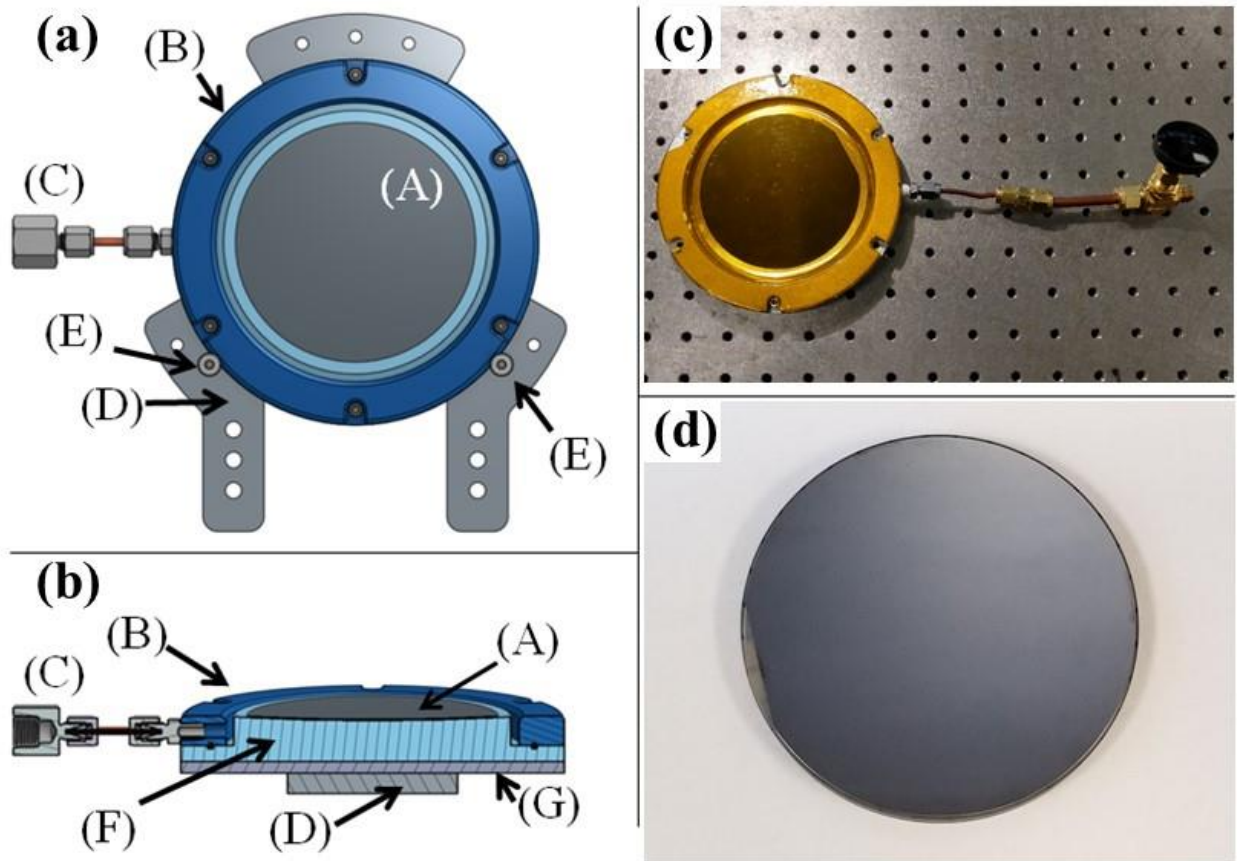

**Figure 1**: **(a)** CAD rendering front view of the vacuum formed bent crystal analyzer (VF-BCA). (A) bent wafer; (B) front flange with polyimide film (not shown); (C) pumping line; (D) aluminum alloy vertical support plate; (E) support bolts to define the position of the outer diameter of the VF-BCA body. **(b)** CAD rendering section view of the VF-BCA. (F) CNC-machined vacuum form; (G) steel backing plate for magnetic mounting, where magnets (not shown) are attached to part (D). **(c)** Photograph of the VF-BCA, note the flexible orange polyimide film that allows air pressure to force the wafer into the shape of the form machined into part (F). A second, similar VF-BCA instead has a simple recess in part (F) to accept a 1-m radius of curvature concave glass lens. **(d)** Photograph of a commercial, anodically-bonded 10-cm diameter SBCA with 1-m radius of curvature.

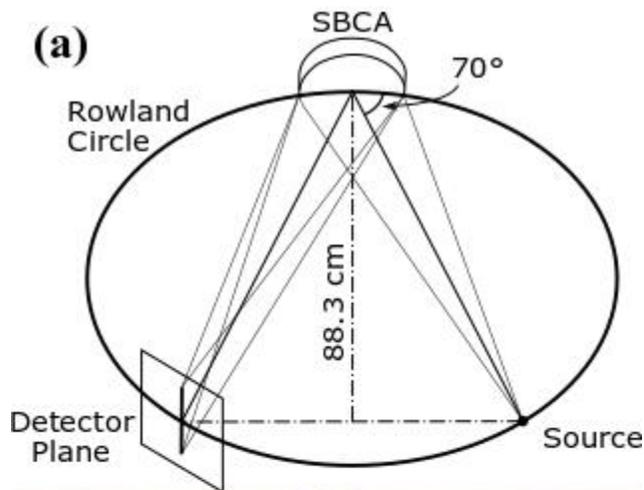

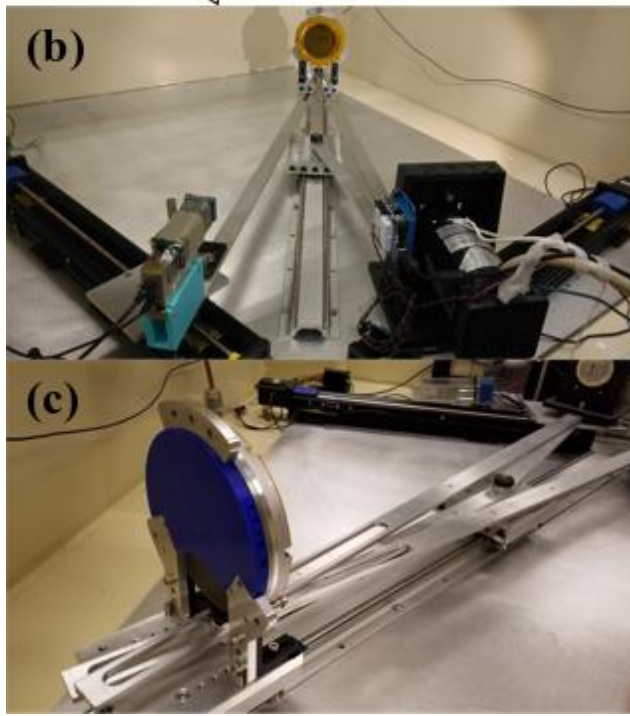

**Figure 2**: **(a)** The Rowland circle implementation for the laboratory spectrometer, shown for a 70 ° Bragg angle. Note the definition of the detector plane. Also note that perfect point-to-point focusing by the optic would require that its radius of curvature out of the Rowland plane be equal to the normal distance from the source-detector arc to the optic, i.e., the sagitta of the reflexive arc from the source point on the Rowland circle to the detector point, indicated as 88.3cm on the diagram above. This motivates the use of toroidally curved forms, as discussed in the text. **(b)** Photograph of the laboratory spectrometer. **(c)** Photograph from the reverse side of a vacuum-formed bent crystal analyzer installed into the spectrometer. Note the use of small magnets inside the plastic 3D-printed part coupled to the steel mounting plate (part (G) in Fig. 1) to hold the analyzer in location but allow easy rotation about the azimuthal angle.

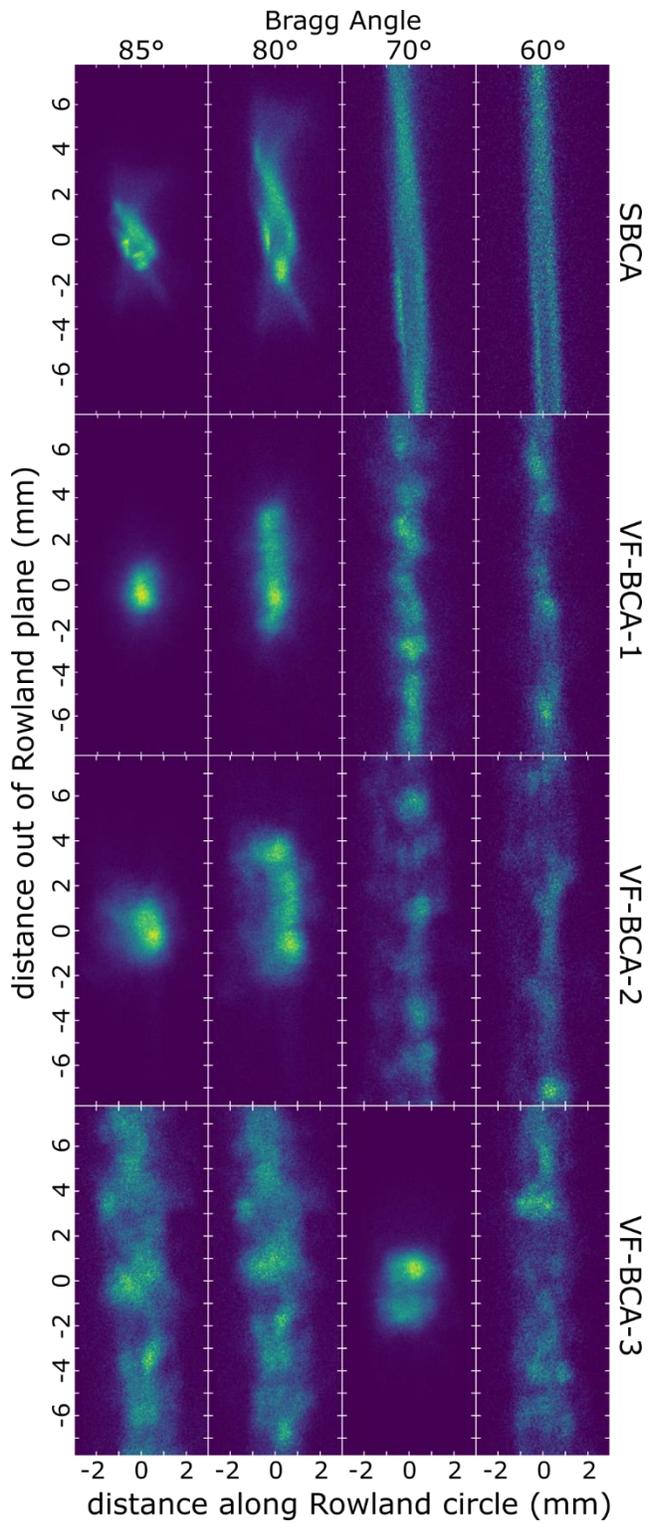

**Figure 3**: Characterization of the spatial distribution of x-ray intensity on the detector plane from **(top)** a commercial Si (551) analyzer (SBCA); **(second from top)** VF-BCA-1 with a Si (711) wafer and **(second from bottom)** VF-BCA-2 with a Si (711) wafer; **(bottom)** VF-BCA-3 with a Si (711) wafer, a torus optimized for $\theta_B = 70°$. The colorscale of each frame is independently normalized; for a comparison of intensities, see Fig. 4.

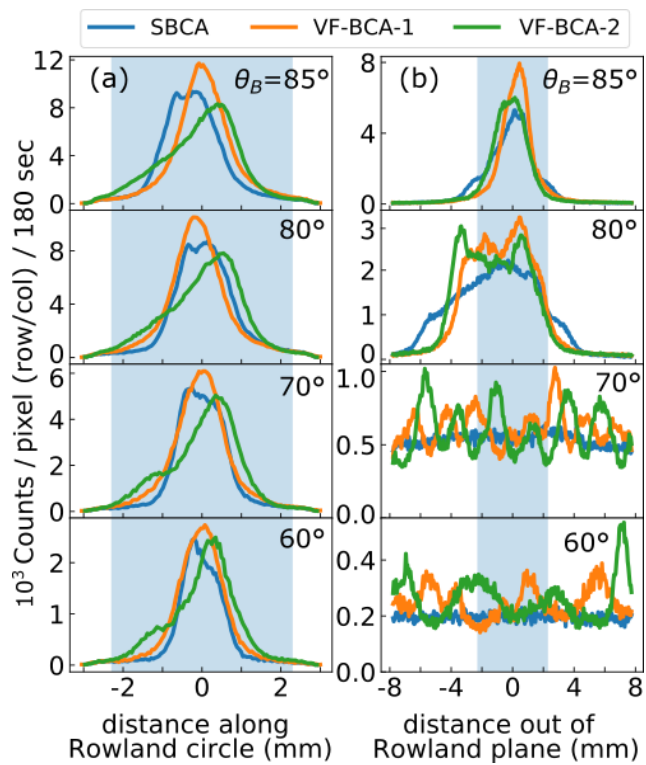

**Figure 4: (a)** Comparison of beam profiles in the Rowland plane for different optics. **(b)** Comparison of beam profiles in the direction perpendicular to the Rowland plane different optics. The extent of the silicon SDD's active region is represented by the shaded regions.

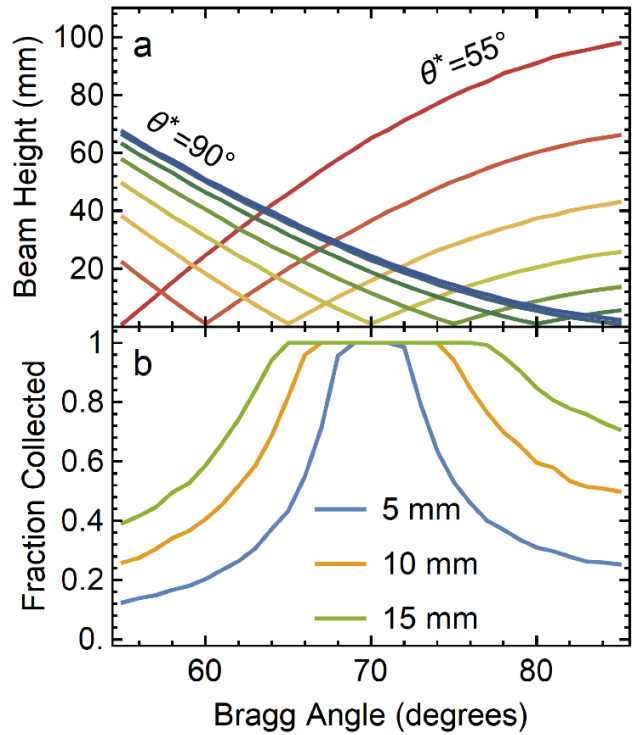

**Figure 5**: **(a)** The modeled beam height as a function of $\theta_B$ for doubly-curved optics with a 1-m principal radius of curvature but with secondary curvatures designed for point-to-point focus as design Bragg angles $\theta^*$ varying in 5° steps from 55° to 90°, the latter being for a fully spherical analyzer. **(b)** The modeled fraction of x-rays hitting three different circular detectors with diameters of 5 mm, 10 mm, and 15 mm.

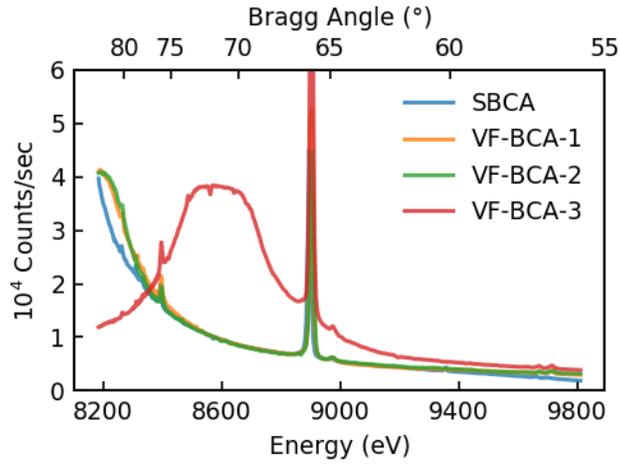

**Figure 6:** Monochromator count rate as a function of photon energy using a Si (551) or equivalent wafer. The sharp features at, e.g., ~8850 eV, are fluorescence lines from the x-ray tube. The rapid roll-off for the spherical optics is due the steadily increasing vertical spread upon decreasing Bragg angle, causing the beam to become taller than the 4-mm active height of the SDD. The toroidal optic shows much improved performance in the designed Bragg angle range.

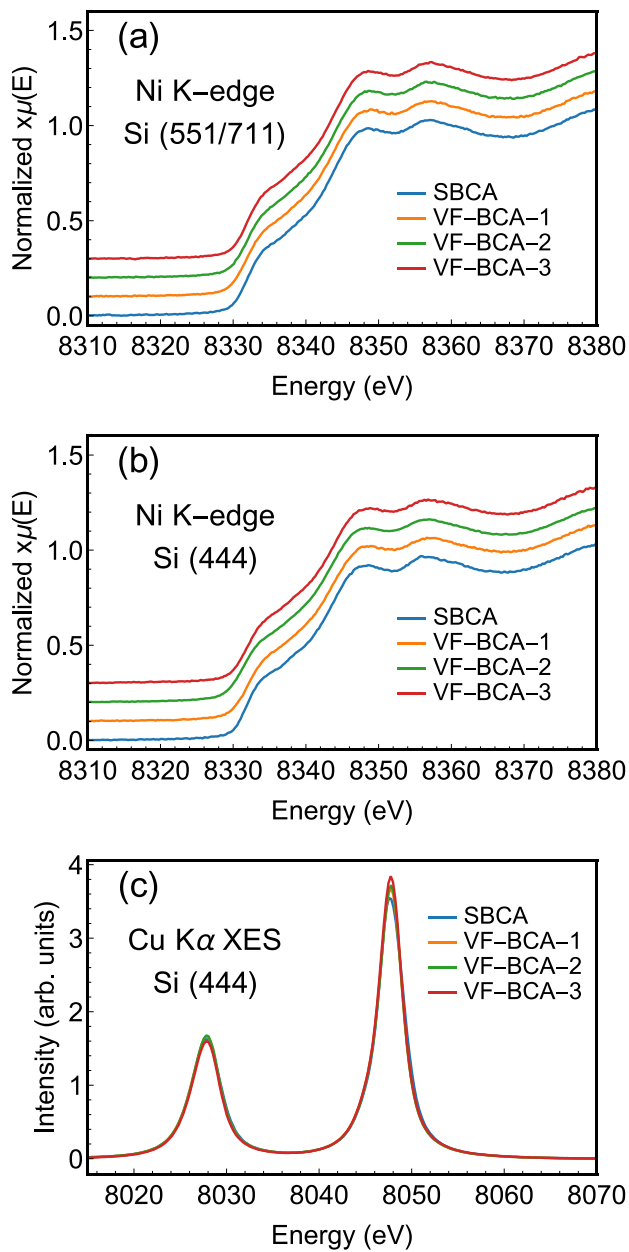

**Figure 7:** Representative spectra demonstrating the capability of the vacuum-clamped optics to perform high-resolution measurements. **(a)** comparison of Ni XANES measured with each optic using Si (551) or equivalent wafers **(b)** comparison of Ni XANES measured with each optic using Si (444) wafers (c) comparison of Cu Kα XES with each optic using Si (444) wafers.